\documentstyle[epsfig,twocolumn,aps,amssymb]{revtex}
\begin{document}
\draft
\title{Vortex line and ring dynamics in trapped Bose-Einstein condensates}
\author{B. Jackson,$^1$ J. F. McCann,$^2$ and C. S. Adams$^1$}
\address{$^1$Department of Physics, Rochester Building, University of Durham, 
South Road, Durham, DH1 3LE, UK.}
\address{$^2$Department of Applied Mathematics and Theoretical Physics, 
Queen's University, Belfast, BT7 1NN, UK.}
\date{\today}
\maketitle
\begin{abstract}
Vortex dynamics in inhomogeneous Bose-Einstein condensates are
studied numerically in two and three dimensions. We simulate the precession of
a single vortex around the center of a trapped condensate, and use the Magnus 
force to estimate the precession frequency. Vortex ring dynamics in a 
spherical trap are
also simulated, and we discover that a ring undergoes oscillatory motion around
a circle of maximum energy. The position of this locus
is calculated as a function of the number of condensed atoms. In the presence 
of dissipation, the amplitude of the oscillation will increase, eventually  
resulting in self-annihilation of the ring.
\end{abstract}
\pacs{PACS numbers: 03.75.Fi, 05.30.Jp, 67.40.Vs} 

\section{Introduction}

Among the most important phenomena associated with Bose-Einstein condensation
(BEC) is the quantization of vorticity, which is intimately connected with
the existence of persistent currents and superfluidity in quantum fluids.
Study of quantized vortices has been confined mainly to liquid HeII
\cite{donnelly}, where detailed comparison to mean-field theory is
complicated by strong interactions between atoms. However, such considerations
are much less important for the recently achieved
BEC in atomic vapors \cite{ande95,davi95,brad95,fried98}. 
In this case the condensate can be accurately described by the 
Gross-Pitaevskii (GP)
equation, an example of a nonlinear Schr\"{o}dinger equation, whose properties
are well-known. This equation admits
vortex solutions, where a non-zero circulation is accompanied by a zero in the
condensate density. The density variation defines the vortex core, with a size
of $\sim 1\, \mu{\rm m}$ ({\it cf.} $\sim 1\, \rm{\AA}$ in HeII). 
Thus, vortices may be directly observable by absorption imaging
\cite{ande95,lund98} or other detection schemes
\cite{gold99,bold99,zamb98,svid98_1}. Apart from their intrinsic interest,
vortices play an important role in the breakdown of superflow in Bose fluids 
\cite{donnelly,fris92,wini99}.

Despite the enormous progress of experiments on atomic condensates 
\cite{kett99}, a vortex state has yet to be observed. One possible procedure 
involves rotating the condensate \cite{marz98,stri98} at a rate
exceeding a critical angular velocity \cite{baym96,sinha97,lund97}, creating
a singly-quantized vortex line along the axis of rotation 
\cite{stri96,dalf96}. Larger angular velocities might be expected to produce 
vortices with higher quanta of circulation. However, previous studies 
\cite{pu98,garc99} suggest that such vortices are
unstable, leading to arrays of singly-charged vortices \cite{butt99,feder99_2}
similar to those found in HeII \cite{donnelly}. 

A separate but complementary idea is to use a tightly-focussed far-off
resonant laser beam \cite{jack98}, which creates vortex pairs when dragged 
through the condensate. Alternatively, by stirring the condensate angular
momentum is transferred, and a single vortex may be produced \cite{cara99}. 
A vortex ring may be formed by translating one condensate through another 
\cite{jack99}, or by 3D soliton decay \cite{jones86,joss95}. 

Experimentally, the Bose-condensed gas is usually confined in a magnetic
trap, often modelled by a harmonic potential. This profoundly alters
the condensate properties in many ways \cite{baym96,dalf99}, the most
apparent of which is its spatial inhomogeneity. This results in a 
variation of
vortex energy as a function of position, with a maximum at
the center of a non-rotating trap. As a
consequence, a single vortex precesses around the condensate center
\cite{bold99,svid98_2}. In addition, the vortex
is thermodynamically unstable, and dissipation at finite temperatures leads to
its expulsion from the cloud \cite{rokh97,fedi99}. The instability is also
apparent in the excitation
spectrum, with the existence of a mode possessing negative energy with respect
to the ground state \cite{garc99,dodd97,isos98}.

In this paper, we study the motion of vortices in trapped Bose condensates, by 
numerical solution of the GP equation. As discussed in Section II, 
this equation is valid in the 
limit of low temperatures, and describes the conservative motion of 
a vortex. Section III presents measurements of the precession 
frequency of a single vortex in two and three dimensions, and compares 
the results to analytical expressions. We use a Magnus force
argument to estimate the precession frequency in 2D.
In Section IV we study vortex rings, and find that they perform a cyclical 
motion. A 
stationary state can be found which corresponds to a ring with maximum energy.
This point of unstable equilibrium is 
equivalent to that of a single vortex in the center of a 
non-rotating condensate. Finally, we summarize in Section V, and briefly 
discuss finite temperature effects.       

\section{Theory}

\subsection{The Gross-Pitaevskii equation}

In experiments on atomic vapours \cite{ande95,davi95,brad95}, evaporative 
cooling can be 
extended to very low effective temperatures, such that the non-condensate
fraction is very small. The densities are sufficiently low that
interactions
can be represented by a pseudopotential of the form 
$U_0 \delta(\mbox{\boldmath $r$}-\mbox{\boldmath $r'$})$, where 
$U_0=4\pi\hbar^2 a/m$, and $a$ is the $s$-wave scattering length. This leads
to the Gross-Pitaevskii (GP) equation \cite{pita61,gross63} for the condensate
wavefunction, $\Psi(\mbox{\boldmath $r$,t})$, given by:      
\begin{equation}
 i \hbar \frac{\partial \Psi}{\partial t} = \left(- \frac{\hbar^{2}}{2m} 
 \nabla^2 + V + N U_0 |\Psi|^2 \right) \Psi,
\label{eq:Gross-P}
\end{equation} 
where $N$ is the number of atoms, each of mass $m$. Consequently, 
$N|\Psi|^2$ is
the condensate number density, which satisfies the normalization condition:
\begin{equation}
 \int |\Psi(\mbox{\boldmath $r$,t})|^2 {\rm d}^3 \mbox{\boldmath $r$} = 1.
\label{eq:Norm}
\end{equation}
The harmonic trapping potential is denoted by 
$V(\mbox{\boldmath $r$})=m(\omega_x^2 x^2+\omega_y^2 y^2
+\omega_z^2 z^2)/2$.

It is convenient to scale (\ref{eq:Gross-P}) in dimensionless units.
We use harmonic oscillator units (h.o.u.) \cite{rupr95}, where the units of
length, time and energy are $(\hbar/2m\omega_x)^{1/2}$, $\omega_x^{-1}$ and
$\hbar \omega_x$ respectively. We also consider a frame rotating about the
$z$-axis with angular velocity $\Omega$, where the angular momentum operator 
is given by $L_z=i(y \partial_x - x \partial_y)$. Retaining the normalization 
condition (\ref{eq:Norm}), Eq.\ (\ref{eq:Gross-P}) becomes:
\begin{equation}
 i \partial_t \Psi=\left(-\nabla^2+\tilde{V}+
 C|\Psi|^2 - \Omega L_z \right) \Psi,
\label{eq:GP-scale}
\end{equation}
where $\tilde{V}=\frac{1}{4}(x^2+\eta y^2+\epsilon z^2)$ and 
the anisotropy parameters are defined as $\eta=\omega_y^2/\omega_x^2$
and $\epsilon=\omega_z^2/\omega_x^2$. The interaction parameter is given by
$C=(NU_0/\hbar\omega_x)(2m\omega_x/\hbar)^{\alpha/2}$, where $\alpha$ is the
number of dimensions. In 2D, where atoms are confined in the
$x-y$ plane, $N$ represents the number of atoms per unit length along $z$.

Eq.\ (\ref{eq:GP-scale}) can be integrated by various numerical methods
\cite{feit82}. As in our previous work \cite{jack98,jack99}, we utilize a Fast
Fourier Transform (FFT) technique. Stationary solutions of (\ref{eq:GP-scale}),
can be represented by
$\Psi(\mbox{\boldmath $r$},t)=\Psi(\mbox{\boldmath $r$}) {\rm e}^{-i\mu t}$ 
where $\mu$ is the chemical potential. They are found by propagating a trial 
wavefunction (e.g.\ a Gaussian solution for $C=0$) in imaginary time, 
$t\rightarrow -i\tilde{t}$.
In contrast to real time, the resulting evolution operator is non-unitary, so 
the
wavefunction must be renormalized after each time-step. The ratio
of norms provides a convenient estimate for the chemical potential,
$\mu=(2\Delta t)^{-1} \ln [\langle|\Psi(t)|^2\rangle/\langle
|\Psi(t+\Delta t)|^2 \rangle]$. In imaginary time, excitations are 
exponentially damped, and both
$\Psi$ and $\mu$ rapidly converge to a stationary solution,
providing accurate initial conditions for time-dependent simulations.

The free energy per particle, $E$, is given by:
\begin{equation}
 E = \int \left( |\nabla \Psi|^2+\tilde{V}|\Psi|^2+\frac{C}{2}|\Psi|^4
 -\Omega \Psi^{*} L_z \Psi \right) {\rm d}^3 \mbox{\boldmath $r$},
\label{eq:E-fnctl}
\end{equation}
where for a stationary solution it can be shown that:
\begin{equation}
 E=\mu-\frac{C}{2} \int |\Psi|^4 {\rm d}^3 \mbox{\boldmath $r$}. 
\label{eq:Emu}
\end{equation}
Imaginary time propagation minimizes the chemical potential, $\mu$.

\subsection{The Vortex State}

Using the Madelung transformation \cite{donnelly},
the condensate wavefunction can be represented in terms of its density
$\rho(\mbox{\boldmath $r$},t)=|\Psi(\mbox{\boldmath $r$},t)|^2$ and phase 
$S(\mbox{\boldmath $r$,t})$
by $\Psi=\sqrt{\rho} \, {\rm e}^{iS}$. The superfluid velocity is given by
${\boldmath v}_s=\hbar(\Psi^{\ast}\nabla\Psi-\Psi\nabla\Psi^{\ast})/im\rho=
(\hbar/m)\nabla S$.
So, the circulation around an arbitrary closed loop is:
\begin{equation}
 \kappa=\frac{\hbar}{m} \oint_{C} \nabla S\cdot d{\boldmath l} = \frac{nh}{m},
\label{eq:Circ}
\end{equation}
where the property of a single-valued wavefunction restricts $n$ to an integer
value. From (\ref{eq:Circ}),
$\nabla\times\mbox{\boldmath $v$}_s=\sum_i 2\pi n_i \delta
(\mbox{\boldmath $r$}-\mbox{\boldmath $r$}_{0i})$, and the superfluid is
irrotational everywhere except for vortices at
$\mbox{\boldmath $r$}=\mbox{\boldmath $r$}_{0i}$. We define a circulation
vector $\mbox{\boldmath $\kappa$}$ on the vortex axis, which has magnitude
$|\mbox{\boldmath $\kappa$}|=\kappa$ and direction
$\nabla\times\mbox{\boldmath $v$}_s$.

A stationary state exists for a single vortex line at
$\mbox{\boldmath $r$}_0=0$. If, for simplicity, we assume a non-rotating 
isotropic trap in
the $x-y$ plane, then this state can be represented as
$\Psi(r,\phi,z)=\sqrt{\rho(r,z)} {\rm e}^{in\phi}$. Substitution into
(\ref{eq:Gross-P}) gives:
\begin{eqnarray}
 \frac{\partial^2 f}{\partial \tilde{r}^2} + \frac{1}{\tilde{r}}
 \frac{\partial f}{\partial \tilde{r}} + \frac{\partial^2 f}
 {\partial\tilde{z}^2}-\frac{n^2 f}{\tilde{r}^2} -
 \frac{\hbar^2\omega_x^2}{4\mu}(\tilde{r}^2+\epsilon \tilde{z}^2)f \nonumber \\
  + f - f^3 = 0,
\label{eq:Vorteqn}
\end{eqnarray}
where $\tilde{r}=(2m\mu)^{1/2} r/\hbar$, $\tilde{z}=(2m\mu)^{1/2} z/\hbar$,
and $f=(NU_0\rho/\mu)^{1/2}=(\rho/\rho_0)^{1/2}$. 
Eq.\ (\ref{eq:Vorteqn}) yields the following asymptotic forms:
\begin{equation}
 \rho \asymp \rho_0 \left(\frac{r}{\xi} \right)^{2n^2}, \qquad r \ll \xi,
\label{eq:Vortdens1}
\end{equation}
\begin{eqnarray}
 \rho \asymp \rho_0 \left(1-\left(\frac{n\xi}{r} \right)^2 -
 \left(\frac{r}{R_{\bot}} \right)^2 - \left(\frac{z}{R_z} \right)^2 
 \right), \nonumber \\
  \xi \ll r < R_{\bot},
\label{eq:Vortdens2}
\end{eqnarray}
and $\rho=0$ for $r \ge R_{\bot}$ and $|z| \ge R_z$, where
$R_{\bot}^2=2\mu/m\omega_x^2$ and $R_z^2=2\mu/m\omega_z^2$. The parameter
$\xi=\hbar/(2m\mu)^{1/2}=(8\pi\rho_0 a)^{-1/2}$ is the coherence
length in the center of the condensate \cite{baym96}, and determines the size
of the vortex core. Eq.\ (\ref{eq:Vortdens2}) is presented in
{\cite{sinha97,stri96,rokh97} and is analogous to the well-known Thomas-Fermi
(TF) approximation for the ground state, valid for large $NU_0$.
In this limit, $\rho_0$ corresponds to
the density in the center of the ground state condensate.

For the special case $n=1$, Eqs.\ (\ref{eq:Vortdens1})
and (\ref{eq:Vortdens2}) can be interpolated by a simple function, which
generalized to a straight, off-axis vortex line at $\mbox{\boldmath $r$}_0$ 
is:
\begin{equation}
 \rho \approx \rho_0 \left(\frac{|\mbox{\boldmath $r$}-
 \mbox{\boldmath $r$}_0|^2}{|\mbox{\boldmath $r$}-\mbox{\boldmath $r$}_0|^2 + 
 \xi^2} \right) \left(1 - 
 \left(\frac{r}{R_{\bot}} \right)^2 - \left(\frac{z}{R_z} \right)^2 \right),
\label{eq:Vortdens3}
\end{equation}
where $\rho=0$ when the right-hand side is negative.
We compare Eq.\ (\ref{eq:Vortdens3}) to the numerical solution of the GP 
equation in Fig.\ 1. The latter is found by imaginary time propagation,
imposing a phase variation of $2\pi$ around the vortex line at each 
time-step.
Eq.\ (\ref{eq:Vortdens3}) provides a good estimate especially in the
high-$N$ limit, in a similar way to the TF approximation to the ground state.  
\begin{figure}
\centering\epsfig{file=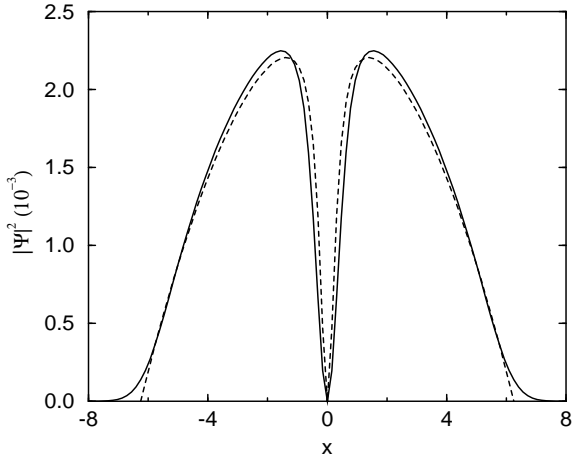, clip=, width=6.25cm, height=8.0cm,
 bbllx=95, bblly=35, bburx=580, bbury=640, angle=270}
\caption{Cross-section through a singly-quantized vortex line, showing 
 condensate density as
 a function of position. The solid line plots the exact 3D wavefunction as 
 calculated from imaginary time propagation of Eq.\ (\ref{eq:GP-scale}), 
 for $C=4000$, $\Omega=0$ and $\epsilon=\eta=1$. The dashed line
 represents the analytic approximation (\ref{eq:Vortdens3}), with 
 $\rho_0 \simeq 2.443 \times 10^{-3}$, $\xi \simeq 0.3199$, $r_0=0$, and 
 $R_{\bot} = R_z \simeq 6.252$. These parameters are calculated from 
 the chemical potential $\mu_{\rm TF}=(15C/64\pi)^{2/5}$, given by the
 normalization condition of the ground-state Thomas-Fermi wavefunction.}
\label{fig:vortstate}
\end{figure}

\section{Single Vortex Motion}

\subsection{Two Dimensions}

In this section we study the dynamics of a singly-quantized ($n=1$) vortex
line in a non-rotating condensate. Imaginary time propagation of the
GP equation (\ref{eq:GP-scale}) is used to provide an initial condition for
the real-time simulation. If we consider a vortex at position
$\mbox{\boldmath $r$}_0$ relative to the axis of a non-rotating trap, then the
free energy of the system, Eq.\ (\ref{eq:E-fnctl}), is found to attain a 
maximum when $\mbox{\boldmath $r$}_0=0$ (see Fig.\ \ref{fig:rotframe-E}). So,
a vortex initially at the
origin will remain stationary, but is unstable to infinitesimal displacements.
The GP equation implies that the system
is Hamiltonian, and therefore an off-axis vortex will follow a path of
constant energy corresponding to precession around the trap center. The
presence of dissipation will lead to drift towards lower energies, causing
the vortex to spiral out of the condensate \cite{rokh97,fedi99}.

If the trap is rotated with angular velocity, $\Omega$, then the energy of
a central vortex, $E_{\rm rot}$, decreases such that 
$E_{\rm rot}=E_n-n\Omega$ \cite{garc99}.
Appearance of a vortex becomes energetically favorable when 
$E_{\rm rot}<E_0$, so that the critical angular velocity is simply:
\begin{equation}
 \Omega_c = \frac{E_n-E_0}{n \hbar}.
\end{equation}
In the TF limit \cite{sinha97,svid98_2,feder99}:
\begin{equation}
 \Omega_c = \frac{5\hbar^2}{8\mu} (\omega_x^2+\omega_y^2)
 \ln \left(\frac{R_{\bot}}{\xi} \right).
\label{eq:crit-vel}
\end{equation}
For $|\Omega| > \Omega_c$, an on-axis vortex attains global stability; 
however, there remains an energy barrier for vortices entering from the edge 
\cite{feder99_2} (see Fig.\ \ref{fig:rotframe-E}).

Inspection of Fig.\ \ref{fig:rotframe-E} also reveals that that above an 
angular velocity 
$\Omega_m$, the vortex attains a local minimum, so a vortex is metastable
when $\Omega_m < |\Omega| < \Omega_c$. In the TF limit,
$\Omega_m=3\Omega_c/5$ \cite{svid98_2}, while in the non-interacting limit
$\Omega_m \rightarrow \Omega_c$ \cite{linn99}. In both the weak and strong 
coupling limits, $\Omega_m \simeq -\omega_a$, where $\omega_a$ is the 
frequency of the so-called `anomalous' mode obtained from solution of the 
Bogoliubov equations \cite{svid98_2,dodd97,isos98,linn99}. It is thought that 
$\omega_a$ corresponds to the frequency of vortex precession, thus linking
vortex dynamics and instability.
\begin{figure}
\centering\epsfig{file=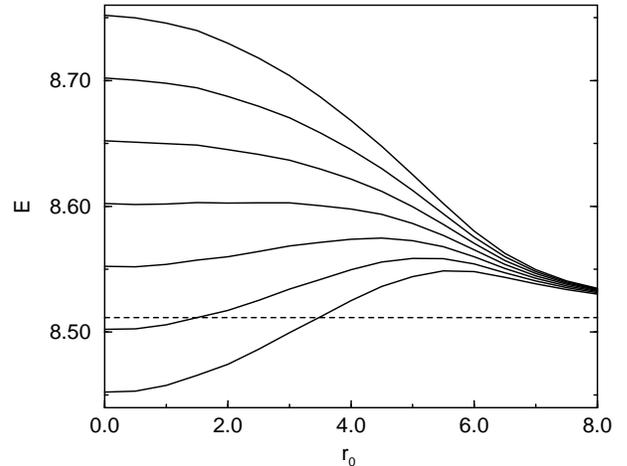, clip=, width=6.25cm, 
 height=8.0cm, bbllx=100, bblly=25, bburx=590, bbury=640, angle=270}
\caption{Energy, $E$, as a function of $n=1$ vortex displacement, in a 2D 
 condensate rotating with angular frequency, $\Omega$ ($C=1000$, $\eta=1$). 
 The top solid curve corresponds 
 to $\Omega=0$, where $\Omega$ increases in steps of 0.05 as one moves towards
 the lowest curve. The dashed line marks the energy of the condensate
 without a vortex.}
\label{fig:rotframe-E} 
\end{figure}
First, we consider the simplest case of a vortex in two dimensions, which
corresponds experimentally to a condensate confined in an axisymmetric
cylindrical trap (where $\epsilon \rightarrow 0$, $\eta=1$). For
$\Omega=0$, simulations show that an off-center vortex accelerates from its
initial condition, soon attaining a near-constant angular velocity $\omega$
around the trap center, such that the instantaneous velocity is
 $ \mbox{\boldmath $v$}_L = \omega \mbox{\boldmath $\hat{\kappa}$} \times 
 \mbox{\boldmath $r$}$.  
The angular velocity, $\omega$, is plotted as a function
of interaction strength and initial position in Fig.\ \ref{fig:prec-C}.
For small $C$, $\omega$ is averaged over a few cycles (e.g.\ 3 revolutions
for $C=200$). However, for higher $C$, numerical instabilities can restrict
the simulations to less than a half-cycle: the error bars reflect the
resulting uncertainty.
\begin{figure}
\centering\epsfig{file=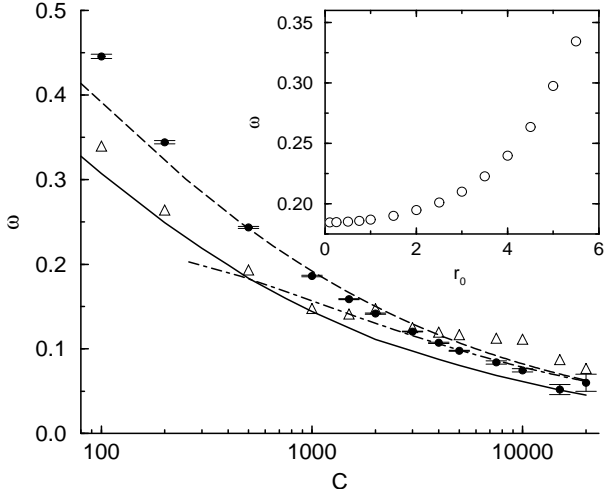, width=6.50cm, height=8.0cm,
 bbllx=95, bblly=45, bburx=580, bbury=630, angle=270}
\caption{Vortex precession frequency, $\omega$, in a 2D condensate
 at $r_0=0.5$,
 as a function of interaction parameter $C=8\pi Na$. Filled circles show
 the results of numerical simulations. 
 The triangles indicate the Magnus force estimates, obtained 
 from the gradient of the numerical values of $E_0$. 
 The analytical Magnus force estimate, Eq.\ (\ref{eq:prec-Mag}), and
 Eq.\ (\ref{eq:prec-Fet}) are
 plotted with dot-dashed and dashed lines, respectively. The TF vortex 
 metastability frequency, $\frac{3}{5} \Delta E$, is plotted as a 
 solid line.
 Inset: $\omega$ as a function of vortex position $r_0$, for
 $C=1000$.}
\label{fig:prec-C}
\end{figure}
The precession frequency decreases as a function 
of increasing interaction strength, $C$. An intuitive semi-analytical argument
for this behavior 
can be formulated in terms of the Magnus effect, familiar from classical
hydrodynamics, and in superfluids and superconductors \cite{sonin97,ao93}.
When the background fluid flows past the circulating fluid connected with the
vortex, a pressure imbalance is created perpendicular to the direction of
the background flow. The resulting Magnus force must
balance the force due to the variation of energy $E$ with position, i.e.\
$\partial E / \partial r_0 = 
m \rho \mbox{\boldmath $\kappa$}\times\mbox{\boldmath $v$}_L$,
where $\mbox{\boldmath $v$}_L$ is the velocity of the vortex line relative to
the ambient condensate. Note that the Magnus force can also be produced by
flow of the condensate around a stationary vortex, as in a rotating
trap. So, one expects that $\omega \simeq \Omega_m$. 

We find $E$ by evaluating the functional, Eq.\ (\ref{eq:E-fnctl}), with
a wavefunction grown in imaginary time with $\Omega=0$. Using the Madelung
transformation, the first term splits into a `quantum pressure' and a 
`kinetic energy' term: 
$|\nabla \Psi|^2 = (\nabla \sqrt{\rho})^2 + \rho (\nabla S)^2$. A numerical
differentiation of $E$ with respect to $r_0$  gives an estimate for $\omega$
using the Magnus force argument (plotted as triangles in 
Fig.\ \ref{fig:prec-C}). However, these estimates are sensitive to small
numerical errors in the energy.

To obtain an analytical estimate, we observe that $\mu$ is approximately 
constant at small $r_0$ and high $C$. This may be
shown by using the decomposition $\Psi(x,y)=\Phi(x,y)\Theta(x,y)$, where an 
off-set vortex core ($\Phi$) is imprinted on the TF ground-state ($\Theta$), 
as implied by Eq.\ (\ref{eq:Vortdens3}). Starting from the GP equation 
(\ref{eq:Gross-P}), and noting that in the TF limit $\Theta$ is slowly 
varying, then it is simple to show that:
\begin{equation}
 - \frac{\hbar^{2}}{2m} 
 \nabla^2 \Phi + \mu_{\rm TF} |\Phi|^2 \Phi = \mu \Phi,
\label{eq:vortmu}
\end{equation} 
where $\mu_{\rm TF}$ is the chemical potential of $\Theta$. As the Laplacian is
spatially invariant, it follows that $\mu$ is independent of the offset at
small $r_0$. This can also be justified numerically, though the approximation
only becomes valid at high $C$. Using Eq.\ (\ref{eq:Emu}) it follows that:
\begin{equation}
 \frac{\partial E}{\partial r_0} \approx -\frac{C}{2} 
 \frac{\partial}{\partial r_0} \left[\int \rho^2 {\rm d}^2 
 \mbox{\boldmath $r$} \right].
\label{eq:Eint} 
\end{equation}
Substituting Eq.\ (\ref{eq:Vortdens3}) for $\rho$ then gives an estimate for
the precession frequency (to logarithmic accuracy):
\begin{equation}
 |\omega| \approx \frac{\hbar \omega_x^2}{\mu} \left[\ln \left(
 \frac{R_{\bot}}{\xi} \right) - \frac{5}{4} \right].
\label{eq:prec-Mag}
\end{equation}

This result may be compared to the expression obtained by Svidzinsky and 
Fetter \cite {svid98_2}, using a time-dependent variational analysis:
\begin{equation}
 |\omega| = \frac{3\hbar\omega_x^2}{4\mu} \ln \left(\frac{R_{\bot}}{\xi}
 \right),
\label{eq:prec-Fet}
\end{equation}
again valid for small $r_0$. These expressions are
plotted together with the numerical results in Fig.\ \ref{fig:prec-C}. In 
addition, we plot $\Omega_m=\frac{3}{5} \Delta E$, where $\Delta E$ is the 
energy difference between the ground and $n=1$ vortex states. Recall that one 
expects that $\omega \simeq
\Omega_m$ in the TF limit. The numerical results lie between
the results of (\ref{eq:prec-Fet}) and the metastability curve. All
of the curves reproduce the correct functional dependence at high $C$. Note 
that these expressions are only
valid for small $r_0$; the vortex precesses faster as it nears the edge of
the condensate, as shown in Fig.\ \ref{fig:prec-C} (inset).  

Compressibility effects become important when the vortex is accelerating or 
when the velocity is an appreciable fraction of the speed of the sound, 
$c_s=(\rho NU_0/m)^{1/2}$. In an infinite 
compressible fluid, phonons may be emitted by a moving vortex, leading to
a drift of the vortex to lower energies \cite{ovch98}. However in a finite
condensate, where excitations remain confined in the region of the vortex,
no net drift is expected. At the beginning of the simulations, we
observe an increase in radius of precession, together with excitation of an 
elliptical center of mass mode
at the trap frequency. In addition, surface waves are created when the vortex
is near the condensate edge. However, as expected we do not observe a
sustained vortex drift (for $T$ up to 110, corresponding to $\sim 6$ full 
cycles for
$C=200$). A drift to lower energies would be expected where a thermal cloud
damps the motion (i.e.\ at finite temperatures). Nevertheless, the vortex 
lifetime is expected to be
long, especially for large numbers of atoms \cite{fedi99}.

\subsection{Three Dimensions}

Vortex dynamics become more complex in 3D, as the vortex line can deform
along its length. In classical and quantum fluids, this can result in the 
propagation of helical waves along the line---so-called Kelvin modes 
\cite{donnelly,pita61}. In simulations of 3D vortex motion, we have observed 
line deformation and oscillations. However, the inhomogeneity of the condensate
complicates matters, and the motion is difficult to resolve into 
simple Kelvin waves characteristic of the bulk condensate. The amplitude of 
the oscillations are typically small, and
as a consequence helical waves are likely to be difficult to detect 
experimentally. Moreover, it is worth noting that the energy of the vortex 
increases as it lengthens. Hence, in the presence of the dissipation the line 
will tend to straighten, effectively damping the Kelvin modes.

Fig.\ \ref{fig:prec3D-C} compares numerically measured values of the 
precessional frequency with the TF result of Svidzinsky and Fetter 
Eq.\ (\ref{eq:prec-Fet}) \cite{svid98_2} . It can be seen that the frequency 
dependence 
is well described; however, the numerical results are significantly
higher ($\sim 20 \%$), but converge slowly to the analytical expression 
towards
high $C$. This disparity may be due to effects resulting from the curvature 
of the line, which will modify predictions that assume a rigid line motion. 
Numerical values of the TF metastability frequency, $\frac{3}{5}
\Delta E$, are also found to be lower than the observed precession 
frequencies.  
\begin{figure}
\centering\epsfig{file=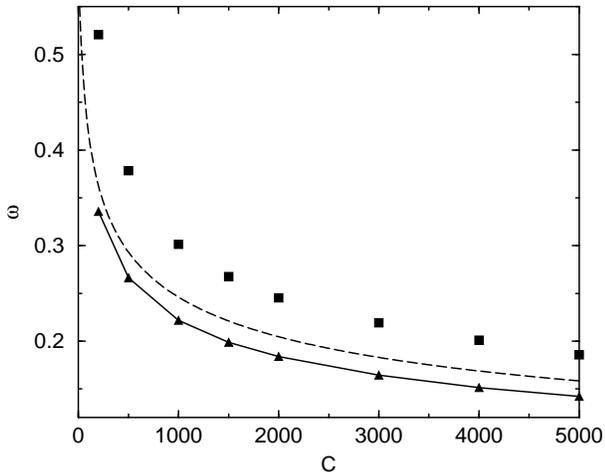, clip=, width=6.25cm, height=8.0cm,
 bbllx=100, bblly=45, bburx=580, bbury=650, angle=270}
\caption{Vortex line frequency, $\omega$, in a 3D condensate 
 ($\eta=1$, $\epsilon=9$)
 at $r_0=0.5$, plotted as a function of interaction parameter, 
 $C=8 \pi Na(2m \omega/ \hbar)^{1/2}$. Squares display numerical results, 
 while the 
 dashed line plots Eq.\ (\ref{eq:prec-Fet}). The solid line shows the TF
 metastability frequency, $\frac{3}{5} \Delta E$.}
\label{fig:prec3D-C}
\end{figure}

\section{Vortex Ring Motion}

The motion of a vortex ring in a trapped BEC may be understood in terms of a
sum of two
contributions to the velocity of each element in the ring. First, 
the precession due to the inhomogeneity of the condensate, as
discussed for a single vortex in Sec.\ III, and second, the velocity
induced by the remainder of the ring, $v_{\rm in}$, which is directed along 
its axis (defined as the $z$-axis). For a spherical condensate,
the total velocity on each element is given by: 
\begin{equation}
 \mbox{\boldmath $v$} = v_{\rm in} \mbox{\boldmath $\hat{z}$} + 
 \omega \mbox{\boldmath $\hat{\kappa}$} \times
 \mbox {\boldmath $r$},
\label{eq:ringvel2}
\end{equation}
where $\mbox{\boldmath $\hat{\kappa}$}$ defines the direction of the 
circulation at the element, and $\omega$ is the precession frequency.
In a homogeneous Bose fluid
$ v_{\rm in} = (\hbar/2mR_r) [ \ln (8R_r/\xi) - 0.615]$ \cite{robe71}, where
$R_r$ is the ring radius. Consider a ring at $z=0$, $r=R_r$.
If the radius is small, the induced velocity dominates
and the ring moves in the $+\mbox{\boldmath $\hat{z}$}$ direction, while
if $R_r$ is large the precession dominates and it travels
backwards. In addition,
the precessional term leads to ring expansion for $z>0$ and contraction
for $z<0$. Thus, the two terms produce an oscillatory motion of the ring.
\begin{figure}
\centering\epsfig{file=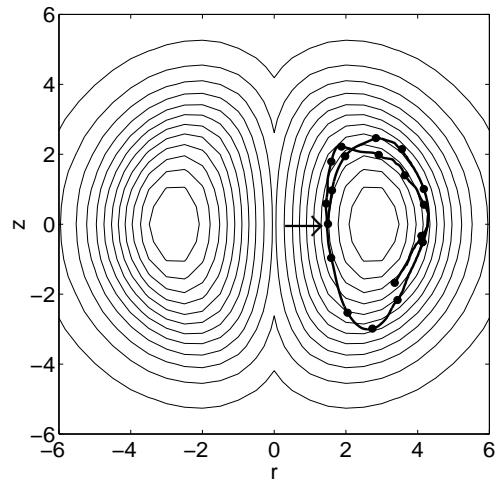, clip=, width=6.5cm, 
 height=6.5cm,
 bbllx=100, bblly=180, bburx=500, bbury=590, angle=0}
\caption{Vortex ring energy in a cylindrically symmetric condensate 
 ($\eta=\epsilon=1$, $C=2000$) as
 a function of radius, $r$, and $z$-position. The energy contours are equally 
 spaced
 between 5.6179 and 6.1159. The bold line shows the motion of one element of
 the vortex ring, where the circles represent the position at equally-spaced
 times (every $T=1$). The ring begins at $(1.5,0)$, marked by an 
 arrow, and cycles around the energy maximum in a clockwise direction.}
\label{fig:ring-E}
\end{figure}
One can also understand the ring motion as a trajectory around an 
energy maximum, in analogy with the single line vortex. To demonstrate 
this, in Fig.\ \ref{fig:ring-E} we plot the energy of an on-axis ring as a
function of its radius, $r$, and $z$-position. Without dissipation 
one would expect the ring motion to follow an energy
contour; however as is apparent in Fig.\ \ref{fig:ring-E} this is not
exactly true. Acceleration of the ring at the beginning of its motion results
in a back-action on the condensate, exciting a center-of-mass mode, and the
subsequent ring dynamics are complicated by the underlying motion of the 
condensate. In
addition, for small $C$ we observe a decay of the ring to lower energies over 
the first cycle of its motion.
As for a single vortex, this effect is associated with the 
compressibility, which results in acoustic emission from the moving
ring \cite{pism93}. 

The energy maximum at $r=R_{\rm eq}$, $z=0$, corresponds to the point where the
two velocity contributions in (\ref{eq:ringvel2}) are equal and opposite, 
leading to a ring in unstable equilibrium. To obtain an 
analytical estimate for this position, we approximate the ring energy by
taking the energy of a single 2D vortex, $E_v$, and integrating around a 
circle 
of radius $R_r$, such that $E_r=2\pi R_r E_v$. The dominant contribution to 
$E_v$ is given by the kinetic energy, so 
$E_v = (m/2) \int \rho v_s^2 {\rm d}^3 \mbox{\boldmath $r$}$. Taking
$\mbox{\boldmath $v$}_s =\mbox{\boldmath $\kappa$} \times
\mbox{\boldmath $r$} / 2\pi r^2$, where we translate the cylindrical 
coordinate system so that
the origin lies on the vortex axis, and using (\ref{eq:Vortdens3}) gives:
\begin{eqnarray}
 E_r \approx \frac{2 \pi^2 \hbar^2 \rho_0 R_r}{m} \Bigg{[} 
 \left( 1- \frac{r_0^2}
 {R_{\bot}^2} \right) \ln \left( \frac {\sqrt{R_{\bot}^2-r_0^2}}{\xi} 
 \right) \nonumber \\
 + \frac{r_0^2}{R_{\bot}^2} - \frac{1}{2} \Bigg{]},
\label{eq:ring-Eapp}
\end{eqnarray}
neglecting terms of order $\xi^2$ and higher. For a ring of radius $R_r$ at 
$z_0$, $r_0^2=R_r^2+z_0^2$. This
expression describes the qualitative features of Fig.\ \ref{fig:ring-E}; 
however, it tends to over-estimate the energy near to the peak ($\sim 10 \%$ at
$C=2000$) and is a poor approximation as $R_r \rightarrow 0$. Nevertheless,
we can obtain an estimate for the equilibrium position from Eq.\
(\ref{eq:ring-Eapp}), which yields $z_0=0$ and:
\begin{equation}
 R_{\rm eq} \approx R_{\bot} \sqrt{ \frac{\ln (\beta R_{\bot}^4) -1}
 {3\ln (\beta R_{\bot}^4)-4} },
\label{eq:ringeqm-app}
\end{equation}
where $\beta=(m\omega/\hbar)^2$. In the TF limit ($R_{\bot} \rightarrow 
\infty$), $R_{\rm eq} \approx R_{\bot} / \sqrt{3} \simeq 0.577 R_{\bot}$, 
which is close to the results from numerical solution, where
$R_{\rm eq} \sim 0.54 R_{\bot}$ at high $C$ (Fig.\ \ref{fig:ringeqm}).
\begin{figure}
\centering\epsfig{file=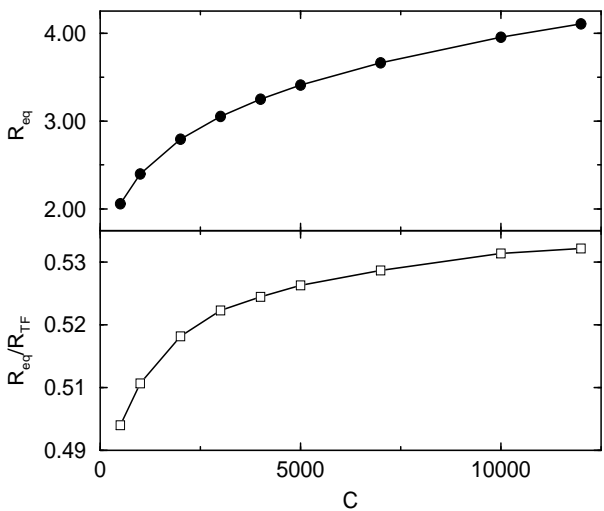, clip=, width=6.75cm, height=8.0cm,
 bbllx=95, bblly=25, bburx=580, bbury=630, angle=270}
\caption{The equilibrium ring radius, $R_{\rm eq}$, as a function of nonlinear
 parameter, $C=8\pi Na (2m\omega/\hbar)^{1/2}$, in a spherically symmetric
 condensate. The lower plot shows the ratio of $R_{\rm eq}$ to the Thomas-Fermi
 radius, $R_{\rm TF} \equiv R_{\bot}=(2\mu_{TF}/m\omega^2)^{1/2}$, where the 
 chemical potential is given by
 the normalization condition of the TF wavefunction, such that 
 $\mu_{\rm TF}=(15C/64\pi)^{2/5}$.}
\label{fig:ringeqm}
\end{figure}
An experimental technique for ring production was 
proposed in Ref.\ \cite{jack99}. The method utilizes a 
two-component BEC, such that when the smaller component, $|2 \rangle$, is 
translated with respect to the other, vortex rings are
created in the larger condensate $|1 \rangle$.

To model the two condensates, a pair of coupled 
GP equations are solved for the wavefunctions 
$\Psi_1 (\mbox{\boldmath $r$},t)$ and $\Psi_2 (\mbox{\boldmath $r$},t)$:
\begin{equation}
 i \partial_t \Psi_i = \left[ -\nabla^{2} + V_i + C  
 \left(|\Psi_i|^2 + |\Psi_j|^2 \right) \right] \Psi_i,
\label{eq:GP-coupled}
\end{equation}
where $i,j=1,2$ $(i \neq j)$ \cite{coup}. The initial 
states of the simulations are created by imaginary time propagation
as previously, where the normalizations are set so that a variable fraction
of atoms are in each condensate, 
$\chi=\langle|\Psi_1|^2\rangle/\langle|\Psi_2|^2\rangle$, and
$\langle|\Psi_1|^2\rangle + \langle|\Psi_2|^2\rangle=1$.
Our simulations then follow the creation and subsequent dynamics of the vortex 
ring, an example of which is shown in Fig.\ \ref{fig:ringmotion}. The
trajectories roughly follow a contour of constant energy 
(see Fig.\ \ref{fig:ring-E}). The ring is created from zero radius
at $t\approx1.6$ and $y\approx1.8$. It then expands and travels forward, before
turning and progressing backwards along the edge of the condensate. Finally, 
it turns again and collapses to a point, where the ring is annihilated. The
annihilation produces a sound wave, which decreases in amplitude as it 
propagates along the 
$z$-direction. The sound wave then disappears at the edge of the cloud.

In the presence of dissipation, annihilation will 
eventually occur for any ring as a culmination of a decay to lower 
energies. This is equivalent to the instability mechanism for a single vortex 
line (Sec.\ III).  

Vortex ring detection is likely to present considerable challenges to
experimentalists. The simplest and most widely used method
of condensate imaging is by measurement of probe laser absorption, after 
release of the condensate from the trap. The subsequent ballistic expansion
results in an effective magnification of a vortex line, which should then be 
visible as a density `hole' \cite{lund98}. However, for rings the density 
minima are obscured
by the rest of the condensate along any line-of-sight. One solution is
to view slices of the condensate after ballistic expansion, using light 
sheets. An alternative method is to study the 
center-of-mass motion of coupled condensates, yielding details of the mutual 
drag that reveal vortex ring formation \cite{jack99}. Collective excitations 
are also utilized in another scheme, proposed in 
\cite{zamb98,svid98_1}, where a vortex line splits the degeneracy
of the quadrupole mode of the condensate. However, further work is needed to
extend this analysis to vortex rings.
\begin{figure}
\centering\epsfig{file=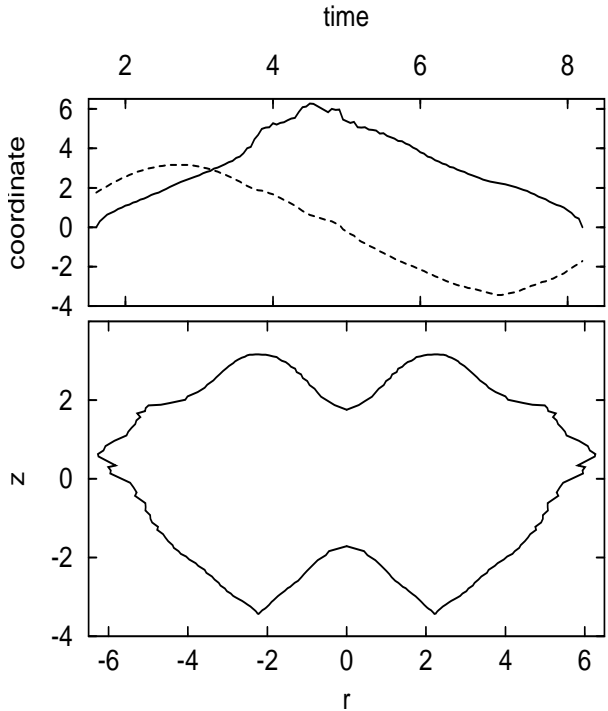, clip=, width=9.5cm, 
 height=8.0cm, bbllx=35, bblly=50, bburx=570, bbury=585, angle=270}
\caption{Vortex ring motion in a condensate ($\eta=\epsilon=1$) after  
 creation from an object. The trajectory of the ring is determined by 
 solving Eq.\ (\ref{eq:GP-coupled}), with $v=1.75$, $C=1100$ and 
 $\chi=10/11$. 
 The upper plot shows the ring radius (solid line) and $z$-coordinate (dashed)
 as a function of time, while a parametric plot (bottom) displays the ring
 radius, $r$, against position $z$.}
\label{fig:ringmotion}
\end{figure}
\section{Summary}

We have studied the motion of vortex lines and rings in
Bose-Einstein condensates in harmonic traps, by numerical
solution of the Gross-Pitaevskii equation. We considered a single vortex in 
two and three dimensions. At the center of a non-rotating condensate the 
vortex state possesses maximum 
energy, corresponding to unstable equilibrium, while an off-center vortex
undergoes precession around this maximum. The precession frequency
was measured and compared to theoretical models.

Vortex rings were also found to undergo an oscillatory motion. For a 
particular radius, the ring energy is a maximum, corresponding to state
of unstable equilibrium. A lower energy ring precesses around the locus of
maximum energy. Also, a ring which is created at a point will eventually 
collapse to a point, resulting in self-annihilation.

The study of dissipative vortex dynamics at finite temperatures is 
particularly interesting. In this case, interaction of the vortex
with the thermal cloud leads to a transfer of energy, and consequently a decay
of the vortex state. One can consider this to be due to anisotropic scattering
of 
thermal excitations \cite{fedi99}, where the momentum transfer is apparent
as a frictional force on the vortex. The transverse component of this force
results
in expulsion of the vortex from the condensate.  A full microscopic model of 
vortex dynamics is needed to describe this behaviour and will form 
the focus of future work.                       

\acknowledgements
Financial support for this work was provided by the EPSRC.


\begin{references}
\bibitem{coup}
 The trap potentials are given by 
 $V_1=(x^2+y^2)/4$ and $V_2=V_1+(x^2+(y-vt)^2)$, where $v$ is a constant
 translational velocity. The scattering lengths between hyperfine states are
 assumed to be equal. Simulation of the two-component dynamics involves 
 propagating (\ref{eq:GP-coupled})
 for $\Psi_1$ and $\Psi_2$ consecutively at each time step. We restrict
 attention
 to the cylindrically symmetric case ($\eta=1$), so that 
 $\nabla^2=(1/r)\partial_r + \partial_{rr} + \partial_{zz}$.
 To propagate (\ref{eq:GP-coupled}), the resulting Hamiltonian is split into
 operators involving the first derivative term, which is dealt with by finite
 differences, and the remainder which is treated by the FFT approach mentioned
 previously. Care must be taken with the singularity at $r=0$, where a forward
 differencing approximation is utilized 
 $(1/r)\partial_r\Psi |_{r=0}=2[\Psi(\Delta r)-\Psi (0)]/(\Delta r)^2 + O
 (\Delta r^2)$, whilst elsewhere central differencing is applied.
\end{references}
\end{document}